\documentclass[prl,aps,twocolumn,superscriptaddress,floatfix]{revtex4-2}  
\usepackage{graphicx}  
\usepackage{dcolumn}   
\usepackage{bm}        
\usepackage{amssymb}   
\usepackage{amsmath}  
\usepackage{color}
\usepackage[utf8]{inputenc}
\usepackage{booktabs} 

\begin{document}

\title{Geometric Constraint on Residue Phases: Resolving the N(2190) Anomaly and Diagnosing Exotic States}

\author{S.~Ceci}
\email{sasa.ceci@irb.hr}
\affiliation{Rudjer Bo\v{s}kovi\'{c} Institute, Bijeni\v{c}ka  54, HR-10000 Zagreb, Croatia}
\author{R.~Omerović} 
\affiliation{University of Tuzla, Urfeta Vejzagića 4, 75000 Tuzla, Bosnia and Herzegovina}
\author{H.~Osmanović} 
\affiliation{University of Tuzla, Urfeta Vejzagića 4, 75000 Tuzla, Bosnia and Herzegovina}
\author{M.~Uroić}
\affiliation{Rudjer Bo\v{s}kovi\'{c} Institute, Bijeni\v{c}ka  54, HR-10000 Zagreb, Croatia}
\author{M.~Vuk\v si\' c}
\affiliation{University of Oxford, Oxford OX1 2JD, United Kingdom}
\author{B.~Zauner}
\affiliation{Institute for Medical Research and Occupational Health, Ksaverska 2, HR-10000 Zagreb, Croatia}
\date{\today}

\begin{abstract}
We derive a parameter-free geometric constraint on residue phases dictated by the pole-threshold angle. Using the N(2190) anomaly as a test case, this constraint reveals a sign ambiguity in prior data; correcting it yields a phase of $-28^\circ\pm10^\circ$, matching our prediction. This consistency validates the method as a model-independent diagnostic for distinguishing compact from molecular states, offering a rigorous tool for exotic spectroscopy.
\end{abstract}

\maketitle

The amplitude of the resonant scattering is dominated by the pole in the complex energy plane \cite{DalitzMoorhouse}. The resonant scattering amplitude near the pole can be described as
\begin{equation}
    T=\frac{|r|\,e^{i\theta}}{M-E-i\,\Gamma/2}+T_B, \label{Eq:Amplitude}
\end{equation}
where $E$ is the energy, $M-i\,\Gamma/2$ is the pole position, $|r|\,e^{i\theta}$ is the pole residue, and $T_B$ is the background term. Close to the pole, Eq.~(\ref{Eq:Amplitude}) with constant $T_B$ provides a good description for the resonant amplitude. Various multi-channel analyses have extracted pole parameters with increasing precision \cite{HohlerPWA,Cutkosky,Batinic1995,Vrana2000,Arndt2006,Batinic2010,WI08,Anisovich2012,Svarc2013,LplusP,Ronchen2015,Sokhoyan2015,Ronchen2022}. The complex pole and the magnitude of the complex pole residue have clear physical meaning, while the physical interpretation of the residue phase remains challenging \cite{PDG}. 

Recent studies suggest a geometric interpretation related to the angle between the pole and the threshold in the complex energy plane \cite{Ceci13,Ceci17}. More generally, there is a mathematical identity connecting the zero, the pole, and the zero of the real part of the amplitude with the residue phase \cite{Ceci26}. For the lightest resonances in a partial wave the elastic threshold plays the effective role of the zero, while the zero of the real part of the amplitude is one of the definitions of the Breit-Wigner mass. Such resonances have relatively small negative residue phases and are classified as Type Ia. If there is strong non-resonant dynamics in the region between the threshold and the pole, the effective zero gets closer to the resonance, and the angle drastically rises, producing a much larger negative residue phase (typically from $-80^\circ$ to $-120^\circ$). These resonances are classified as Ib. Subsequent resonant poles in partial waves can have zeros of the amplitude anywhere around them, producing basically any value of the residue phase, and they are dubbed Type II. 

In Ref.~\cite{Ceci26}, it was shown that the resonant lowest-lying states among $N^*$ and $\Delta$ resonances are either Ia or Ib type with a sole exception: $N(2190) 7/2^-$. The PDG estimate for the phase was an anomalous value of $0^\circ\pm30^\circ$, while the model predicted drastically smaller $-45^\circ\pm23^\circ$. 

In this Letter, we scrutinize both the theoretical constraint—which forbids a zero or positive phase for the isolated lowest-lying resonance in a partial wave—and the partial wave analyses underpinning the PDG estimate: Cutkosky's CMB \cite{Cutkosky}, \v{S}varc's L+P \cite{LplusP}, Sokhoyan's CBELSA/TAPS \cite{Sokhoyan2015}, and R\"onchen's J\"ulich model \cite{Ronchen2022}. Through a detailed examination of the input parameters and the extracted phases, we identify the source of the discrepancy. Rectifying the sign ambiguity changes Sokhoyan's result to $-28^\circ\pm10^\circ$, identical to the revised PDG average. This model prediction also changes by removing the outlier pole mass from the PDG average. With the new result of $-26^\circ\pm9^\circ$, and the amplitude matching the partial wave data, we find the N(2190) anomaly fully resolved. 

While the identification of isolated narrow resonances is often straightforward, the spectroscopy of broad states near open thresholds poses a formidable challenge. This regime is paramount for the modern study of exotic hadrons, such as $XYZ$ states and pentaquarks \cite{Guo2018,LHCb2022}. In these systems, strong background interference and threshold cusps can distort lineshapes, making the extraction of pole parameters highly model-dependent \cite{Mikhasenko2015,Pilloni2017}. The distinction between a genuine resonant pole and a kinematic effect—such as a triangle singularity or a non-resonant cusp—hinges on the phase behavior. 
In contrast, genuine resonant poles dictate a specific phase evolution. Since kinematic singularities generally fail to satisfy the geometric constraint derived here, it serves as a rigorous falsification test and a model-independent diagnostic tool. Thus, resolving the N(2190) anomaly serves as a proof-of-principle for applying this constraint to the exotic sector.

The model used here was proposed in Ref.~\cite{Ceci17}, and refined in Ref.~\cite{Ceci26}. In it, the background is a complex constant that drives total amplitude to simple zero at the threshold energy ($E_0$). Its value is not fixed by that simple demand, because there is an overall phase which is fixed by the position of the zero of the real part of the amplitude (in some definitions, that is the Breit-Wigner mass $M_\mathrm{BW}$). The scattering amplitude becomes
\begin{equation}
T=\frac{\overbrace{x\,\Gamma/2}^{|r|}\, \,\,\overbrace{e^{i(\alpha+\beta)}}^{e^{i\theta}}}{M-E-i\,\Gamma/2}+\overbrace{x\,e^{i\beta}\sin\alpha}^{T_B},\label{eq:our_amplitude}
\end{equation}
where $x$ is the branching fraction, threshold phase $\alpha$ is 
\begin{equation}
    \alpha = -\arctan\, \frac{\Gamma/2}{M-E_0},
\end{equation}
and Manley's phase $\beta$ \cite{Manley} is given by 
\begin{equation}
       \beta = -\arctan\, \frac{M_\mathrm{BW}-M}{\Gamma/2}.
\end{equation}

\begin{figure}[h!]
    \centering
    \includegraphics[width=0.48\textwidth]{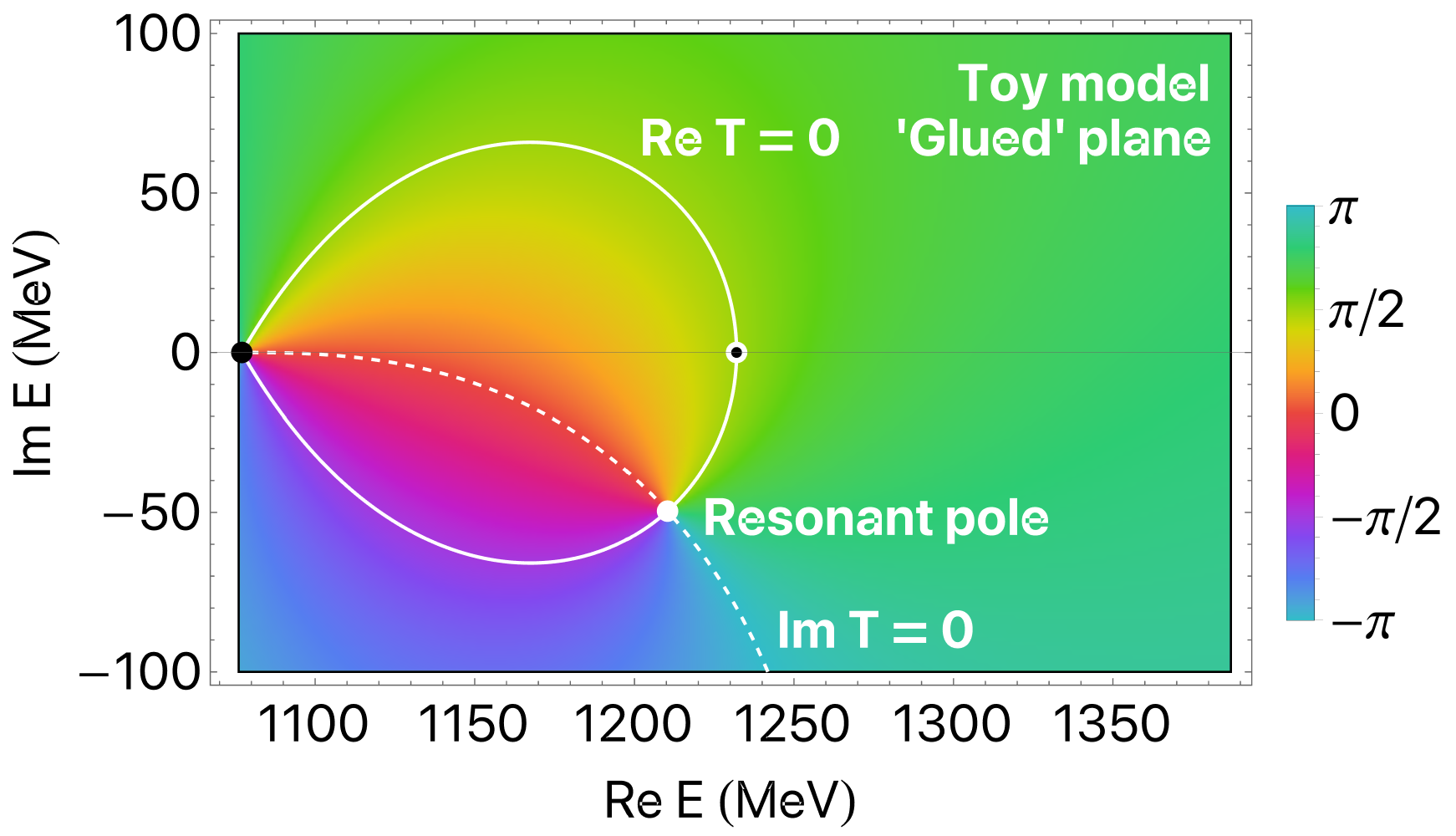}
   \includegraphics[width=0.48\textwidth]{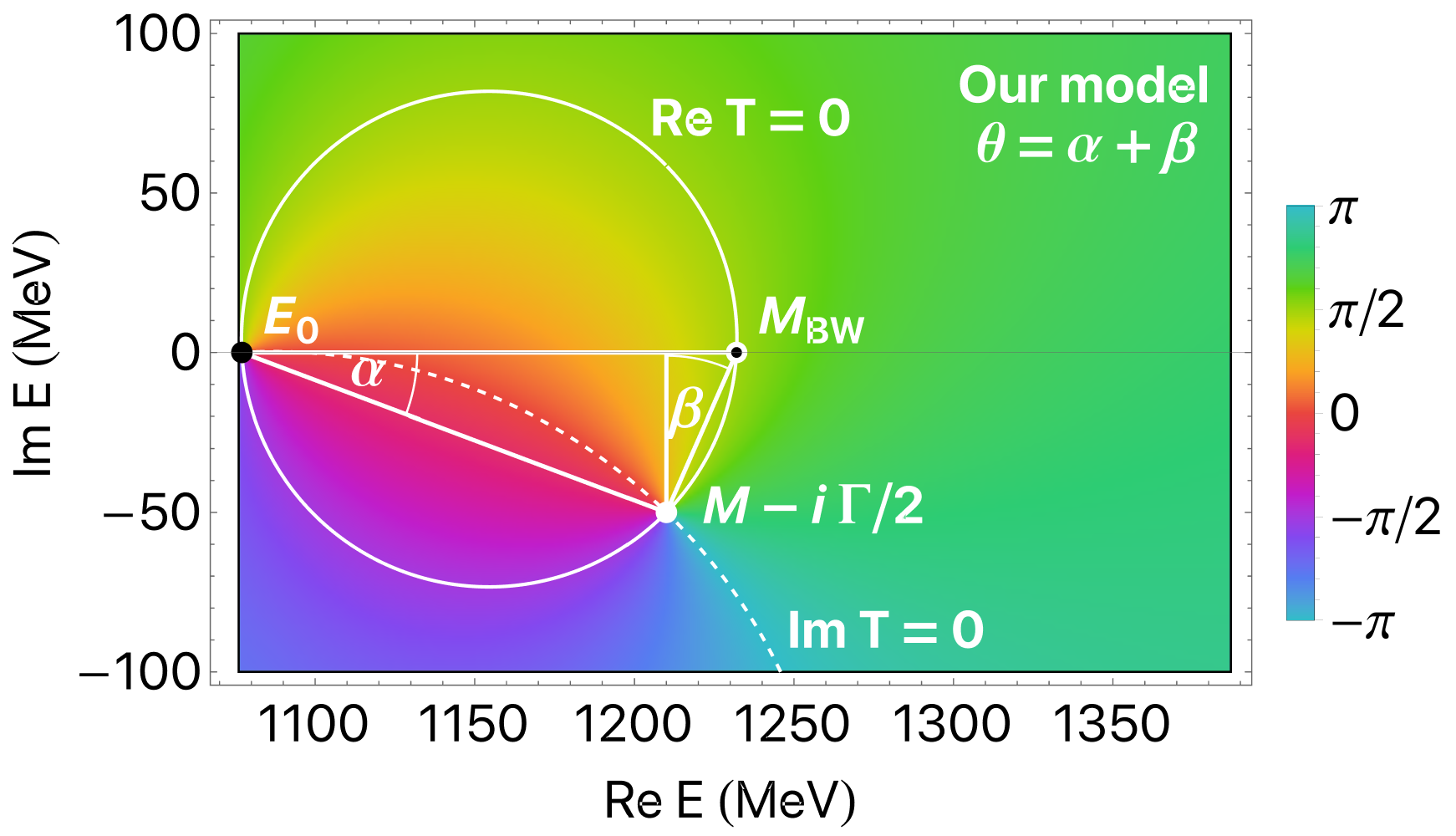}
\caption{Complex phase of the scattering amplitude $T$. (Top) The unitary toy model, connecting the physical (upper half-plane) and non-physical (lower half-plane) Riemann sheets. (Bottom) Our effective model [Eq.~(\ref{eq:our_amplitude})]. Note the remarkable similarity near the pole position ($M-i\Gamma/2$, white circle) and the Breit-Wigner mass ($M_\mathrm{BW}$, bullseye marker). Deviations are visible primarily near the threshold ($E_0$, black circle).}
     \label{fig:Riemann}
\end{figure}

To assess the impact of the non-analytic nature of this model, we compare it to a unitary toy model featuring energy-dependent width with the correct threshold behavior ($q^{2l+1}$). Despite the distinct topology of the unitary toy model (branch cut vs.~simple pole), Fig.~\ref{fig:Riemann} confirms that our effective model captures the relevant analytic structure near the resonance pole.

While the unitarity violation in our effective model is minimal near the pole, we assess its impact by comparing our predictions directly to the single-energy WI08 dataset by the GWU group \cite{WI08}. Our inputs are strictly the PDG parameter estimates; the branching fraction $x$ is taken from the original analysis by Arndt {\it et al.} \cite{Arndt2006}. In our formalism, $x$ acts merely as a real-valued scaling factor and does not influence the residue phase.

There is another important reason we want to compare our amplitudes to the data: the threshold behavior. The true amplitude, as well as the data, must show $q^{(2l+1)}$ behavior near the threshold, while for our amplitude, the threshold is just a simple zero, independent of $l$. Therefore, we examine a selection of $N^*$ and $\Delta$ resonances with increasing orbital angular momentum $l$. 

\begin{figure*}[t]
    \centering
    \includegraphics[width=0.32\textwidth]{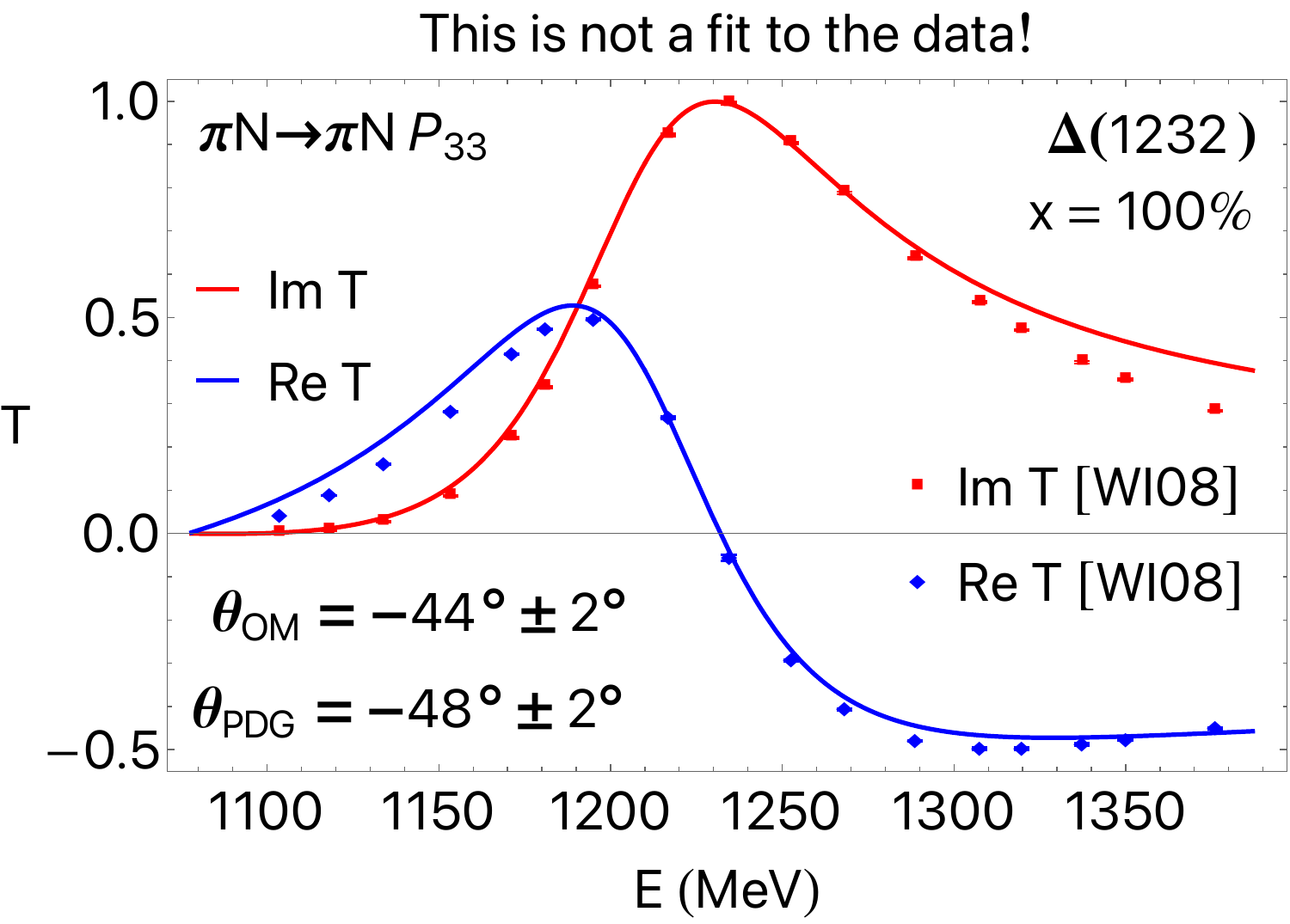}  
    \includegraphics[width=0.32\textwidth]{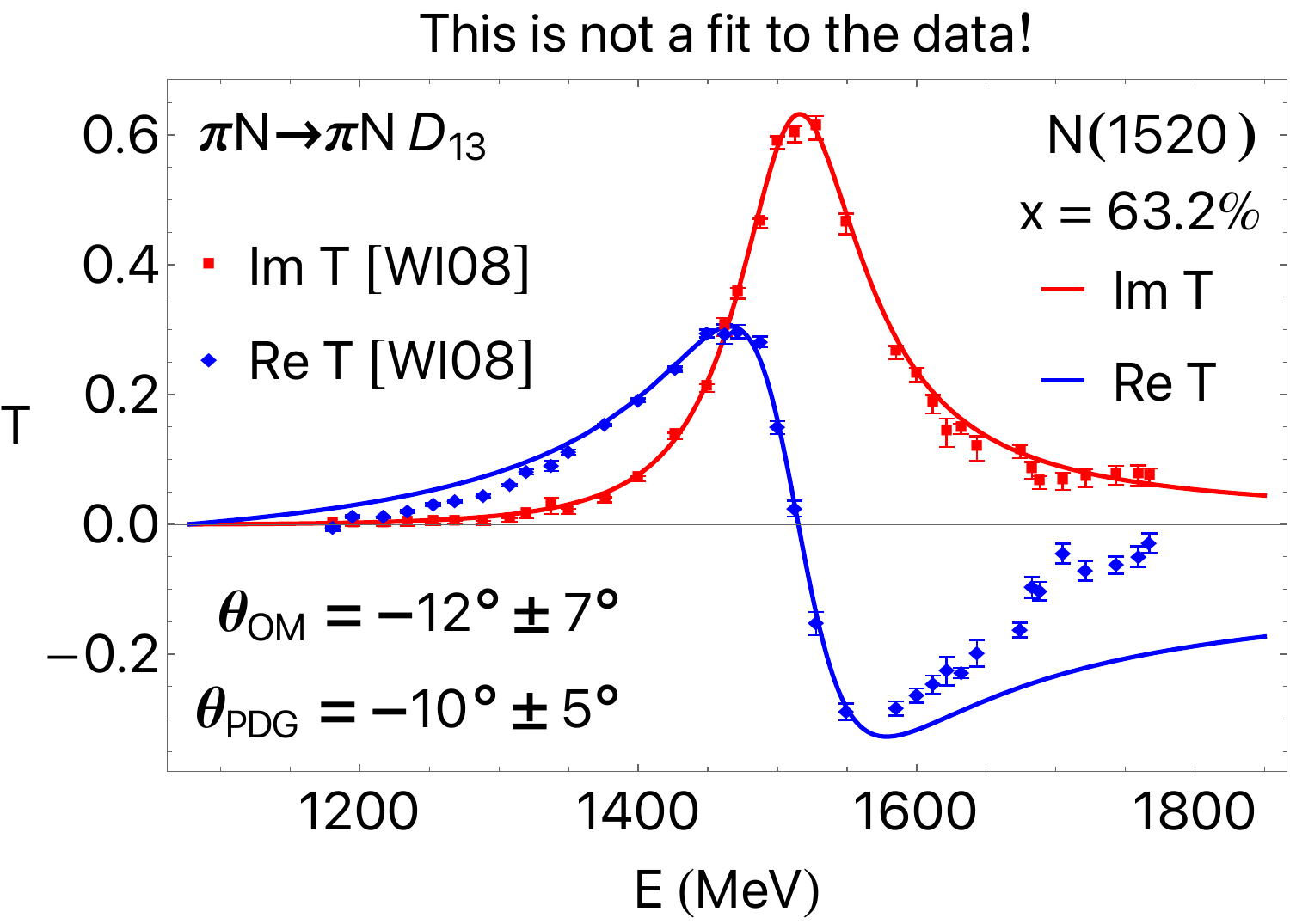}   
    \includegraphics[width=0.32\textwidth]{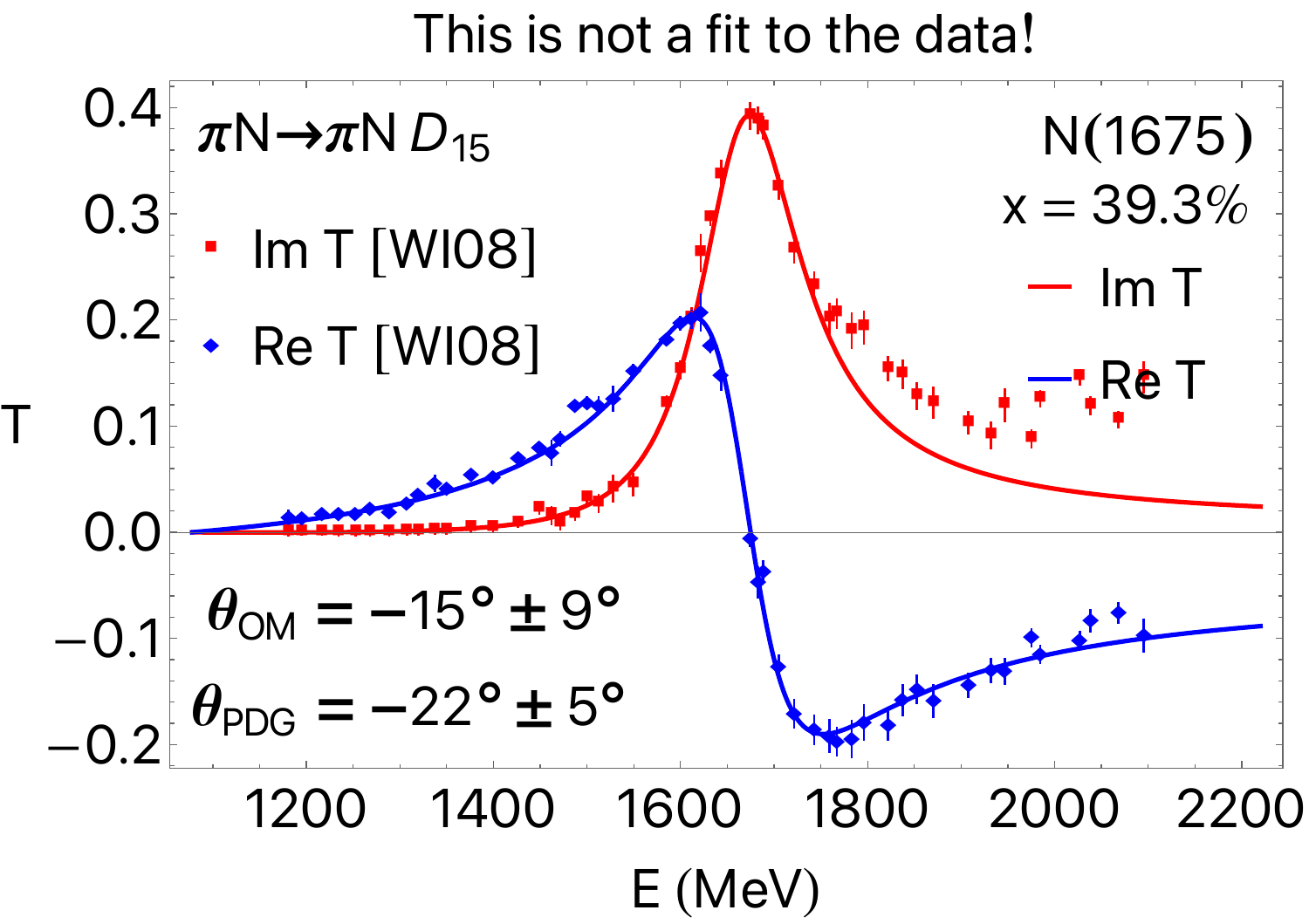}        \includegraphics[width=0.32\textwidth]{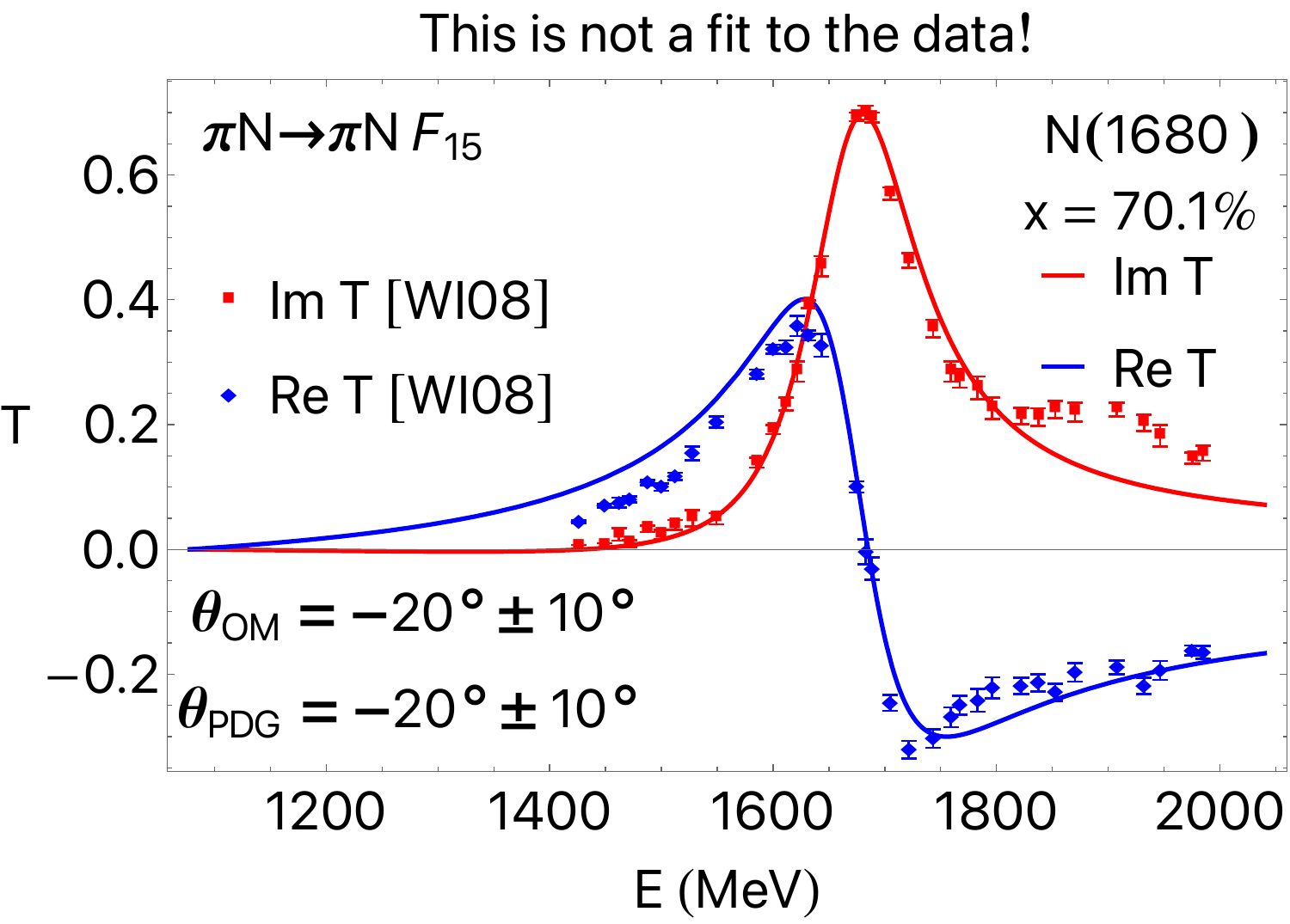}         \includegraphics[width=0.32\textwidth]{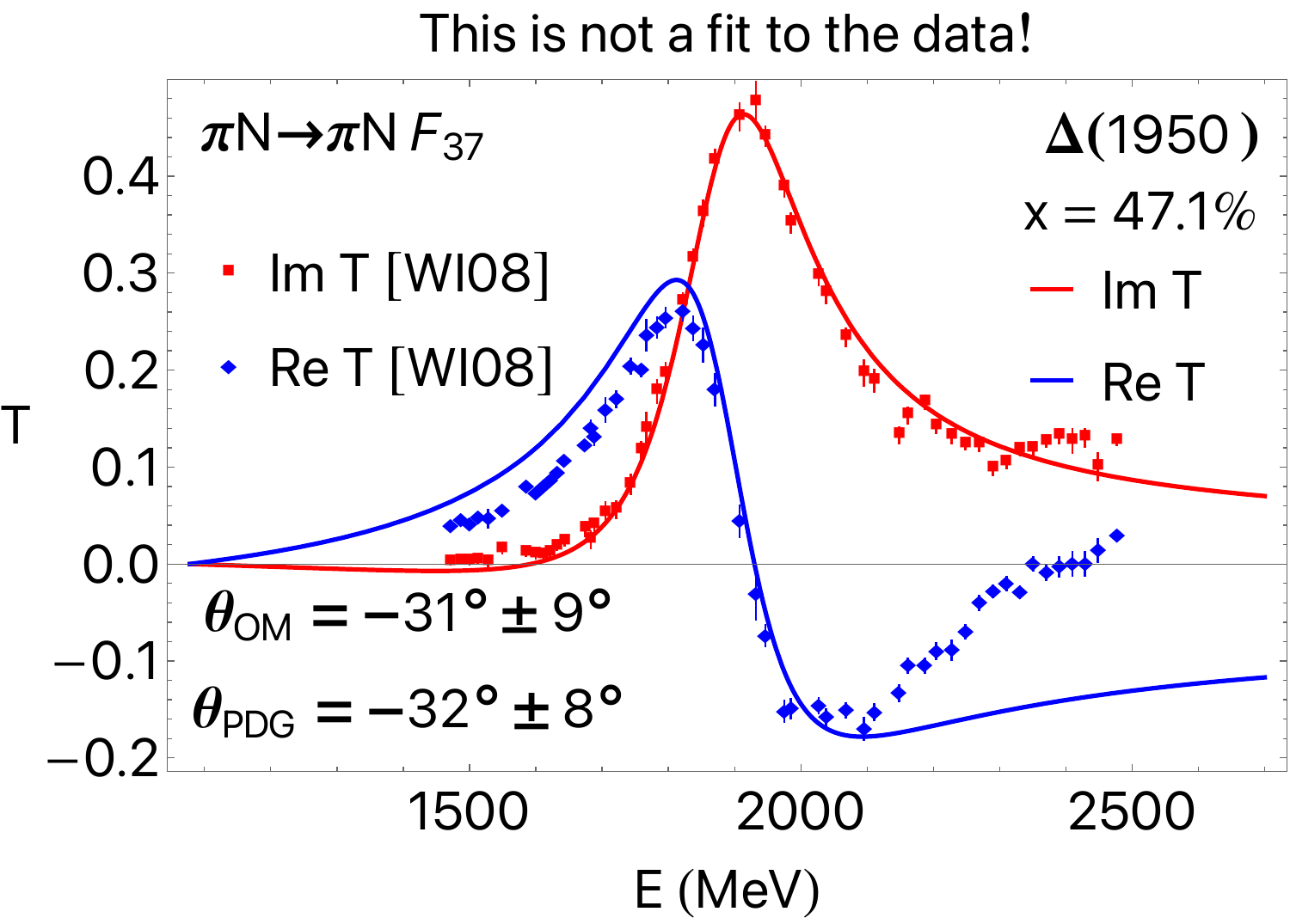}
    \includegraphics[width=0.32\textwidth]{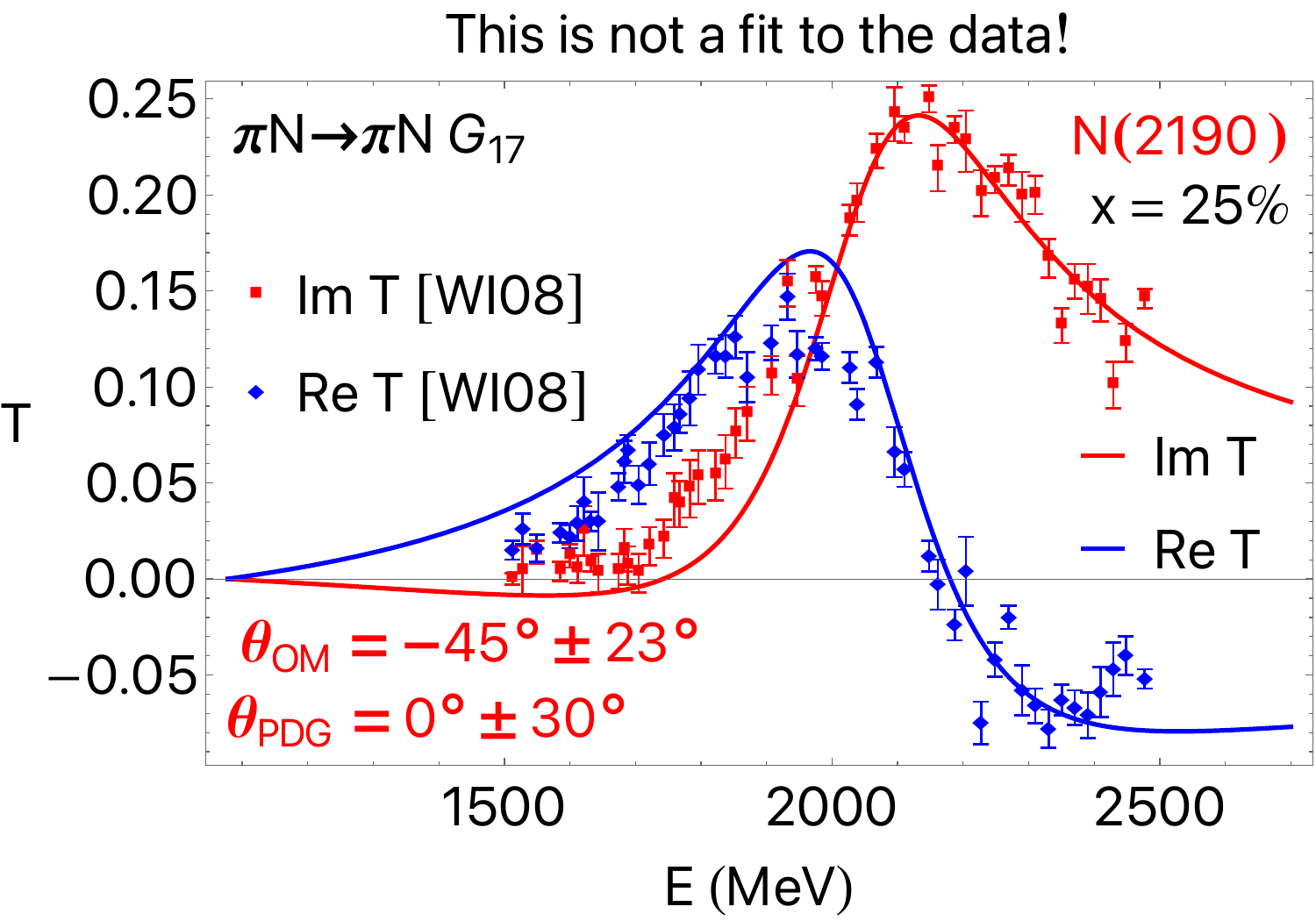}
     \caption{A comparison (not a fit) of our prediction, based on the PDG estimates of masses and widths, to the $\pi N$ elastic WI08 single energy data \cite{WI08}. Evidently, our prediction is not getting worse with higher partial waves. The real part departs somewhat closer to the threshold, but the imaginary part is rather good there. Notably, with exception of N(2190), higher partial waves have increasingly better agreement between our residue phase prediction ($\theta_\mathrm{OM}$) and the PDG estimate ($\theta_\mathrm{PDG}$).}
     \label{fig:threshold}
\end{figure*}

Before showing our results, it is crucial to emphasize that the curves in Fig.~\ref{fig:threshold} are \textbf{not fits to the data}, but zero-parameter predictions generated solely from the PDG masses and widths via Eq.~(\ref{eq:our_amplitude}). The remarkable agreement for the imaginary parts across different partial waves validates the predictive power of the geometric constraint. While the real part exhibits slight deviations near the threshold due to the lack of unitarity corrections, the approximation remains sufficient for robust residue phase extraction. 

For D and F waves, the agreement between our predictions and PDG values is excellent. However, a significant discrepancy appears in the $G_{17}$ partial wave for N(2190). Here, the predicted amplitude exhibits a substantial deviation from the experimental data in both real and imaginary parts, and the extracted residue phase differs from the PDG estimate by $45^\circ$, a tension of approximately $2\sigma$.

We note here that the residue phase $\theta$ is given by the sum of $\alpha$ and $\beta$, as stated in Eq.~(\ref{eq:our_amplitude}), and that we do the error estimate using the standard error propagation formula. Although the uncertainties in mass and width are likely correlated, in the absence of the covariance matrix, we treat them as independent to provide a first-order estimate of the phase uncertainty.

\begin{figure}[h!]
    \centering \includegraphics[width=0.48\textwidth]{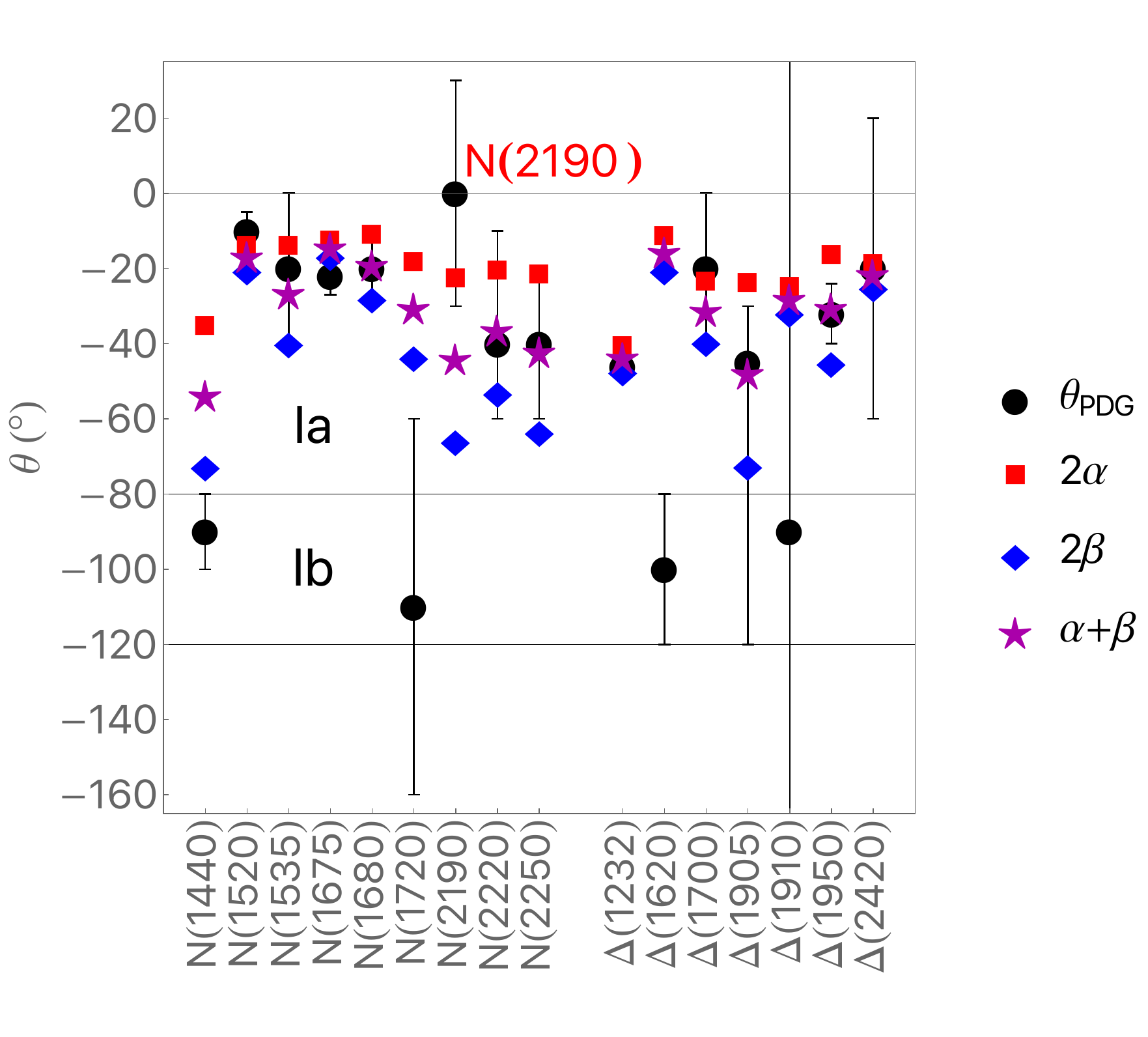}    
    \caption{ The $\pi N$ elastic residue phase for the lightest resonances in partial waves. Black circles are PDG estimates, red squares show the value of twice the threshold angle $\alpha$, blue diamonds show twice the Manley angle $\beta$, and purple stars (the average between the two) are our predictions. Based on agreement with our model, first resonances are classified in types Ia and Ib as introduced in Ref.~\cite{Ceci26}. The N(2190) resonance is the only one not fitting this scheme.}
    \label{fig:AlphaBeta}
    % File: Complex alpha 2024.nb
\end{figure}

Is this discrepancy a natural limit of our model due to the incorrect threshold behavior? To test this, we analyze the lowest-lying $N^*$ and $\Delta$ resonances in various partial waves (Fig.~\ref{fig:AlphaBeta}). The red squares represent the phase derived solely from the threshold angle ($2\alpha$), while blue diamonds represent the phase from the Manley angle ($2\beta$). Our prediction (purple stars) is their average \mbox{($\alpha+\beta$)}. As expected, the more elastic the resonance, the closer these two values converge.

Before we address the issue of N(2190), we want to explain the classification of resonances introduced in Ref.~\cite{Ceci26}. First resonances in partial wave that satisfy our model are classified as Type Ia. Those that have their dominant zero drifted from the threshold closer to the pole (to the left, and slightly down), have larger (more negative) $\alpha$, and also larger $\beta$. Consequently, their $\theta$ gets larger (more negative) and goes into the typical region from $-80^\circ$ to $-120^\circ$. This is what happens to Roper resonance, N(1720), $\Delta(1620)$, and $\Delta(1910)$. These resonances are classified as Type Ib. For them, more general formula is provided in the upper mentioned reference. The Ib classification provides room for resonances that do not satisfy our formula, but N(2190), with its phase of $0^\circ$, is far off that region.

Another important observation clearly visible in Fig.~\ref{fig:AlphaBeta} is that the model-dependent Breit-Wigner mass is crucial for obtaining the correct value of $\pi N$ elastic residue phase in our model (purple stars). The prediction is given by the average of $2\alpha$ and $2\beta$ which are sometime very close to the each other, e.g.~in the case of N(1520), and sometimes rather far away, as in the cases of N(2190) or N(2250). This distance is evidently uncorrelated with the prediction error estimate, and according to Ref.~\cite{Ceci24} it decreases as the resonance becomes increasingly elastic. This strong variability of $\beta$ ($\alpha$ is fixed by the threshold and the pole position) confirms the need for the Breit-Wigner mass. That is highly problematic since it is model-dependent parameter and everyone has their own definition. Nevertheless, it was shown in Ref.~\cite{Ceci26} that mathematically, the proper definition of the Breit-Wigner mass to use in this model is zero of the real part of the amplitude. 

The distinction between the pole position and the Breit-Wigner mass is often subtle but crucial, especially for broad resonances coupled to nearby thresholds. This issue is well-known in the meson sector, for example in the analysis of the $a_1(1260)$ resonance in $\tau$ decays, where dispersive corrections significantly shift the resonance parameters compared to naive Breit-Wigner parameterizations \cite{Mikhasenko2018}. Our model encapsulates these dispersive effects into a compact geometric constraint on the residue phase, providing a transparent link between the pole and the physical scattering amplitude.

Finally, we test our hypothesis about $G_{17}$ wave and N(2190) being a natural limit for our approach. In Fig.~\ref{fig:AlphaBeta} we see that our model predicts almost perfectly phase for the first $H_{19}$ wave resonance N(2220), as well as the $G_{19}$ wave resonance N(2250). Among the $\Delta$ resonances, we see that our model gives also a good predictions for $H_{3\,11}$ resonance $\Delta(2420)$. Evidently, it is neither the threshold issue, nor too high partial wave. 

Now we turn to the building blocks of the PDG estimates, the residue phases provided by the four research groups. In Table \ref{tab:PDG-N2190} are shown their results, as well as parameters needed for our model, and its predictions. The solution by Rönchen {\it et al.}~yields a significantly lower pole mass. While currently an outlier, future determination of a corresponding Breit-Wigner mass for this solution would allow our model to test its internal consistency. The residue phase magnitude extracted by Sokhoyan {\it et al.}~aligns with our prediction almost perfectly, but the sign is opposite. All other lowest-lying resonances follow the Type Ia/Ib classification (either small or large, but definitely negative phases), strongly suggesting a sign convention ambiguity in the extracted solution rather than a physical anomaly.

\begin{table}[h!]
    \centering
    \setlength{\tabcolsep}{2pt} 
\resizebox{\columnwidth}{!}
    {%
    \begin{tabular}{lccccc} 
    \hline\hline
    \textbf{N(2190)} & $M$ & $\Gamma$ & $M_\mathrm{BW}$ & $\theta_\mathrm{OM}$ & $\theta_\mathrm{PDG}$\\
    Source & (MeV) & (MeV) & (MeV) & $(^\circ)$ & $(^\circ)$\\
    \hline
    R\"onchen \cite{Ronchen2022} & $1965^*\pm6$ & $287\pm33$ & N/A & N/A & $-45\pm 14$\\
    Sokhoyan \cite{Sokhoyan2015} & $2150\pm25$ & $325\pm25$ & $2205\pm18$ & $-27\pm 9$ & $28^{\dagger}\pm 10$\\
    Švarc \cite{Svarc2013} & $2079\pm10$ & $509\pm17$ & $2129^\S\pm12$ & $-25\pm 4$ & $-18\pm 3^\ddagger$ \\
    Cutkosky \cite{Cutkosky} & $2100\pm50$ & $400\pm160$ & $2200\pm70$ & $-38\pm 25$ & $-30\pm 50$\\ 
    PDG est \cite{PDG} &   $2050\pm100$ & $400\pm100$ & $2180\pm40$ & $-45\pm23$ & $0\pm30$\\
    \hline
    Arndt \cite{Arndt2006} & $2070$ & $520$ & $2152\pm1$ & $-32$ & $-32$ \\
    \hline 
    $\mathrm{PDG}$ ave &  $2099^*\pm29$ & $425\pm97$ & $2153\pm8$ & $\mathbf{-26\pm9}$ &  $\mathbf{-28\pm10^{\dagger,\ddagger}}$\\
    \hline\hline
    \end{tabular}
    }
    \caption{Input parameters and derived results. Note the significantly lower mass reported by R\"onchen (*), the anomalously small uncertainty in \v{S}varc ($\ddagger$), and the striking agreement in magnitude between the Sokhoyan extraction and our prediction ($\dagger$), differing only in sign. For the \v{S}varc entry, $\theta_\mathrm{OM}$ was calculated using an estimated $M_\mathrm{BW}$ determined from the zero of the real part of the amplitude ($\S$). The Arndt result is included for comparison but excluded from the average due to the absence of uncertainty estimates. The final row presents simple statistical averages of the tabulated parameters; the modified global average for $\theta_\mathrm{PDG}$ incorporates a sign inversion for the Sokhoyan result and conservatively rescales the \v{S}varc uncertainty to match the next most precise measurement ($10^\circ$).}
    \label{tab:PDG-N2190}
\end{table}

We revise the global analysis using simple statistical averages of the PDG parameters, excluding only the low-mass outlier from Ref.~\cite{Ronchen2022}. The residue phase calculated in this manner yields $-26^\circ\pm9^\circ$. To test consistency, we re-evaluate the world average assuming a sign convention mismatch in the result of Sokhoyan {\it et al.}, the only extraction yielding a positive phase. Furthermore, to prevent the anomalously small uncertainty reported by \v{S}varc {\it et al.} from dominating the weighted average, we conservatively rescale their error to match the next most precise measurement ($10^\circ$). Under these conditions, the recalculated world average becomes $-28^\circ\pm10^\circ$. This result is not only in remarkable agreement with our zero-parameter prediction but is also identical in magnitude to the Sokhoyan extraction, differing only by the predicted sign.

\begin{figure}[h!]
    \centering \includegraphics[width=0.475\textwidth]{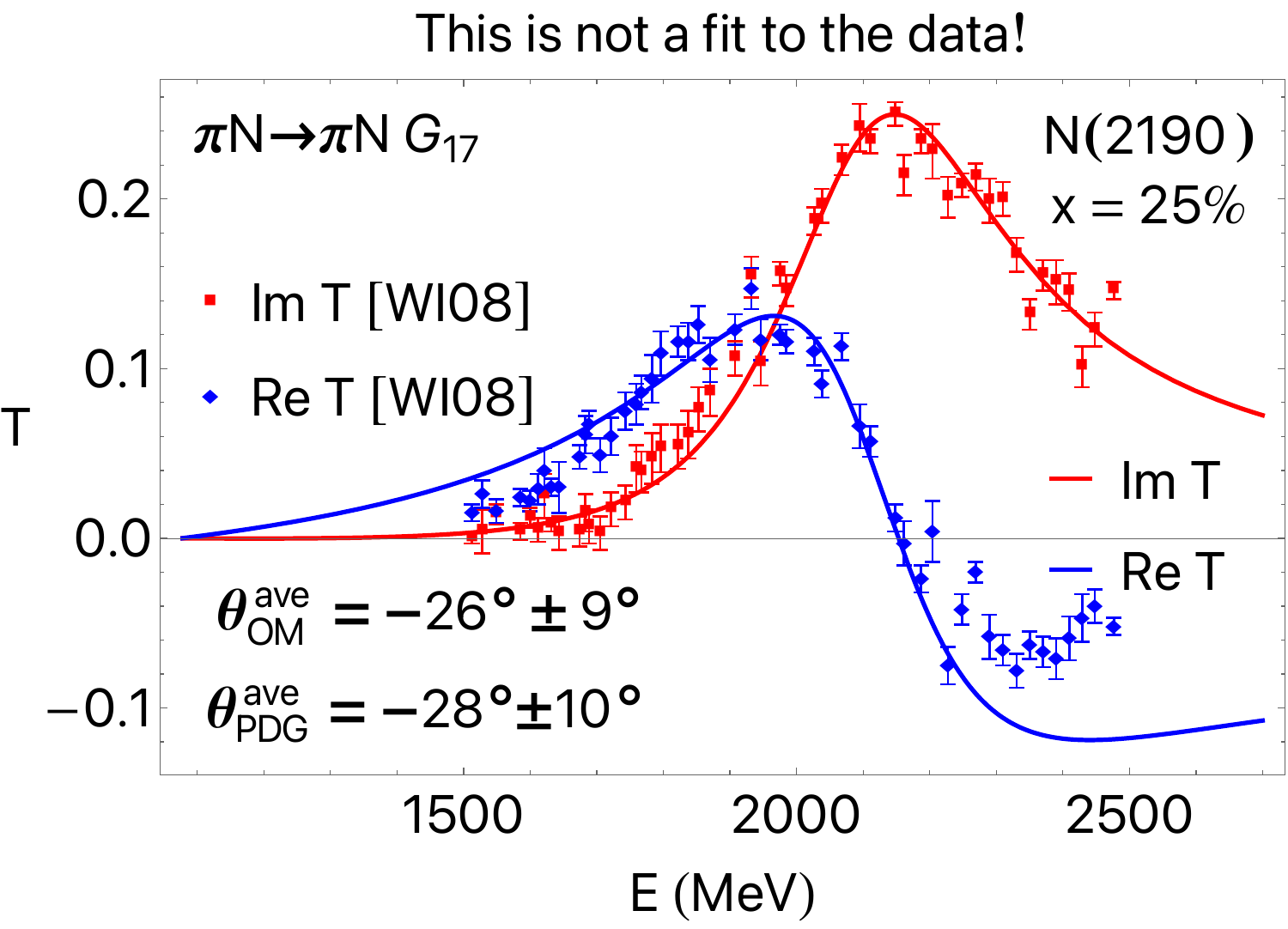}    
    \caption{Resolution of the N(2190) anomaly. Amplitude prediction calculated with the statistical averages of the PDG parameters falls to the data, and the predicted (OM) and measured (PDG) residue phase are fully consistent.}
    \label{fig:N2190-corrected}
\end{figure}

This mechanism extends beyond baryon spectroscopy. In the search for exotic hadrons, such as $XYZ$ states or pentaquarks, distinguishing compact multiquark states from hadronic molecules is of utmost importance. Our model suggests that genuine compact states near threshold should adhere to Type Ia behavior (small negative phases in accordance with geometric constraint). Conversely, loosely bound molecules or states generated by triangle singularities typically force the amplitude zero to drift significantly, resulting in Type Ib behavior (large negative phases, violating the geometric constraint entirely). Thus, the residue phase serves as a topological diagnostic for the internal structure of near-threshold exotics.

In summary, the convergence of the corrected world average and our theoretical prediction strongly supports the Type Ia classification for N(2190), identifying it as a conventional resonant state. The alternative—a positive phase—would imply an unprecedented Type II behavior for a lowest-lying state. This result validates the residue phase as a rigorous, model-independent diagnostic tool. By distinguishing between the geometric signature of compact quark cores (Type Ia) and extended hadronic structures (Type Ib), this constraint offers a vital criterion for disentangling genuine states from kinematic artifacts in the burgeoning sector of exotic spectroscopy.

\end{document}